\documentclass{article}
\usepackage{spconf}

\usepackage[pdftex]{graphicx}
\usepackage{xcolor}
\usepackage[hidelinks]{hyperref} 
\usepackage{microtype}  

\definecolor{Paired-1}{RGB}{31,120,180}
\definecolor{Paired-2}{RGB}{166,206,227}
\definecolor{Paired-3}{RGB}{51,160,44}
\definecolor{Paired-4}{RGB}{178,223,138}
\definecolor{Paired-5}{RGB}{227,26,28}
\definecolor{Paired-6}{RGB}{251,154,153}
\definecolor{Paired-7}{RGB}{255,127,0}
\definecolor{Paired-8}{RGB}{253,191,111}
\definecolor{Paired-9}{RGB}{106,61,154}
\definecolor{Paired-10}{RGB}{202,178,214}
\definecolor{Paired-11}{RGB}{177,89,40}
\definecolor{Paired-12}{RGB}{255,255,153}

\usepackage{amsmath,amssymb,amsfonts,stmaryrd}
\usepackage{bm}
\usepackage{dsfont} 
\usepackage{numprint}
\newcommand{\isep}{\mathrel{{.}\,{.}}\nobreak}

\usepackage{adjustbox}
\usepackage{booktabs}
\usepackage{multirow}

\usepackage[caption=false,font=footnotesize]{subfig}
\usepackage{tikz}
\usetikzlibrary{matrix, positioning, patterns, shapes, arrows}
\usepackage{pgfplots}
\usepgfplotslibrary{groupplots}
\usepgfplotslibrary{colorbrewer}

\usepackage[ruled, linesnumbered, vlined]{algorithm2e}

\newcommand{\furkan}[2]{\ifx&#2&{\leavevmode\color{magenta}#1}\else{\leavevmode\color{magenta}FURKAN\{}#1{\leavevmode\color{magenta}\}}\footnote{{\leavevmode\color{magenta}#2}}\PackageWarning{Furkan}{#1: #2}\fi}

\usepackage[draft]{fixme} 
\fxsetup{theme=color}

\usepackage{fancyhdr}
\AtBeginDocument{\setlength\abovedisplayskip{4pt}}
\AtBeginDocument{\setlength\belowdisplayskip{4pt}}
 
\title{High-Throughput VLSI architecture for Soft-Decision decoding with ORBGRAND}
\name{ \begin{tabular}{c}Syed Mohsin Abbas, Thibaud Tonnellier, Furkan Ercan, Marwan Jalaleddine and Warren J. Gross\end{tabular}}
\address{Department of Electrical and Computer Engineering\\
McGill University, Montr\'eal, Qu\'ebec, Canada}
\fancypagestyle{FirstPage}{

\lfoot{Copyright 2021 IEEE. Published in ICASSP 2021 - 2021 IEEE International Conference on Acoustics, Speech and Signal Processing (ICASSP), scheduled for 6-11 June 2021 in Toronto, Ontario, Canada.}

}
\begin{document}
%
\maketitle
\begin{abstract}
Guessing Random Additive Noise Decoding (GRAND) is a recently proposed approximate Maximum Likelihood (ML) decoding technique that can decode any linear error-correcting block code. Ordered Reliability Bits GRAND (ORBGRAND) is a powerful variant of GRAND, which outperforms the original GRAND technique by generating error patterns in a specific order. Moreover, their simplicity at the algorithm level renders GRAND family a desirable candidate for applications that demand very high throughput. This work reports the first-ever hardware architecture for ORBGRAND, which achieves an average throughput of up to $42.5$ Gbps for a code length of $128$ at an SNR of $10$ dB. Moreover, the proposed hardware can be used to decode any code provided the length and rate constraints. Compared to the state-of-the-art fast dynamic successive cancellation flip decoder (Fast-DSCF) using a 5G polar $(128,105)$ code, the proposed VLSI implementation has $49\times$ more average throughput while maintaining similar decoding performance.
\end{abstract}



\section{Introduction}\label{sec:intro}
Channel coding techniques are an integral part of all modern communications systems. Since their inception \cite{Hamming50}, a lot of effort was focused on finding practical channel coding schemes that could approach channel capacity. Over time, various capacity-approaching codes have been designed, such as Turbo codes \cite{turbo} and LDPC codes \cite{Gallager62}. Polar codes \cite{Arikan09}, proposed in 2009, are able to asymptotically achieve the channel capacity. Each of these aforementioned channel coding techniques, along with many others, require a dedicated decoder. However, there exists an alternate paradigm of decoders that do not rely on the underlying channel code and hence can be used to decode any code.

Guessing Random Additive Noise Decoding (GRAND) is a recently proposed approximate Maximum Likelihood (ML) decoding technique for linear error-correcting codes \cite{Duffy19TIT}. Instead of decoding the received codeword, GRAND attempts to guess the noise present in the codeword. Hence, GRAND can be used for any linear block code. GRAND relies on the generation of putative test error patterns that are successively applied to the received vector and checked for codebook membership using the parity check matrix of the underlying code. The order in which these putative test error patterns are generated is the key difference between different variants of GRAND. Among its variants, GRAND with ABandonment (GRANDAB) \cite{Duffy19TIT} is a hard decision version of GRAND which generates the test error patterns in the increasing Hamming weight order, up to weight $AB$. Symbol Reliability GRAND \cite{SRGRAND} is another variant, which uses thresholding on the input LLRs, to differentiate between reliable and unreliable bits of the received codeword. 

Concomitant with the discovery of the GRAND algorithm, applications that require short frames are emerging, such as machine-to-machine communications. Thus, the implementation of the ORBGRAND algorithm is particularly appealing, since it can approach the ML performance, while only requiring a fraction of its complexity \cite{duffy2020ordered}. ORBGRAND is a soft-input version of GRAND, which generates test error patterns in the logistic weight ($LW$) order rather than in the Hamming weight (HW) order. The logistic weight corresponds to the sum of the indices that are being flipped. The complexity of ORBGRAND is directly dependent on the number of putative test error patterns. ORBGRAND utilizes the soft information associated with the channel observation more efficiently than SRGRAND, which makes it an attractive choice for implementation. In this paper, we investigate parameters that impact the ORBGRAND decoding performance as well as the computational complexity for the polar code studied in \cite{Duffy20205g}. Furthermore, this work presents the first-ever high throughput hardware architecture for ORBGRAND which can achieve an average throughput of $42.5$ Gbps for a code of length $128$ at an SNR of $10$ dB. \thispagestyle{FirstPage}


The remainder of this paper is structured as follows: Section 2 provides an overview of the ORBGRAND algorithm. In Section 3, logistic weights and their impact on complexity and performance are investigated. In Section 4, the proposed hardware architecture is detailed and the implementation results are compared. Finally, concluding remarks are drawn in Section 5.

\section{Background}\label{sec:background}


Algorithm \ref{alg:ORBgrand} depicts the pseudo-code of ORBGRAND. 
The inputs to the algorithm are the vector of channel LLRs $\bm{y}$ of size $n$, a $(n-k)\times n$ parity check matrix of the code $\bm{H}$, a
$n\times k$ matrix $\bm{G}^{-1}$ such that $\bm{G}^{-1}\cdot \bm{G}$ is the $k\times k$ identity matrix, with $\bm{G}$ a 
generator matrix of the code, and the maximum logistic weight considered $LW_\text{max}$. We note that $LW_\text{max} \leq \frac{n(n+1)}{2}$.

First, the indices are sorted in the ascending order of their absolute value LLRs ($\bm{y}$). The obtained permutation vector is denoted by $\bm{ind}$ (line 1). Then, for each logistic weight, all the integer partitions are generated (line 4).
These integer partitions and the permutation vector of the sorted LLRs are used to generate the test error patterns (line 6). Finally, the hard decision obtained from the LLRs ($\hat{\bm{y}}$) is combined with the current test error pattern, and the resulting word is queried for codebook membership (line 7). If the word considered is a codeword, the message is recovered (line 8) and the process is terminated. Otherwise, following error patterns or larger logistic weights are considered.


\begin{algorithm}[t]
\caption{\label{alg:ORBgrand}ORBGRAND Algorithm}
    \DontPrintSemicolon
    \SetAlgoVlined  
    \SetKwData{e}{$\bm{e}$}
    \SetKwData{p}{$\bm{p}$}
    \SetKwData{s}{$\bm{S}$}
    \SetKwData{ind}{$\bm{ind}$}
    \SetKwData{LLR}{$\bm{y}$}
    \SetKwData{sortSet}{$[\bm{r},\bm{ind}]$}
    \SetKwData{estm}{$\hat{\bm{u}}$}
    \SetKwData{ginv}{$\bm{G}^{-1}$}
    \SetKwData{LW}{${LW_\text{max}}$}
    \SetKwData{HW}{$\bm{HW}$}
    \SetKwData{yhat}{$\hat{\bm{y}}$}
    \KwIn{\LLR, $\bm{H}$, \ginv, \LW}
    \KwOut{\estm}
    \SetKwFunction{RecursiveComputeLLRs}{recursiveComputeLLRs}
    \SetKwFunction{DecodeRONE}{decodeR1}
    \SetKwFunction{RDecodeRONE}{redecodeR1}
    \SetKwFunction{DecodeRZERO}{decodeR0}
    \SetKwFunction{Find}{findCandidate}
    \SetKwFunction{new}{generateErrorPattern}
    \SetKwFunction{intPartition}{generateAllIntPartitions}
    \SetKwFunction{Sort}{sortIndices}
    $\ind \leftarrow$ \Sort{\LLR}\;
    $\e \leftarrow \bm{0}$\;
    \For{$i \gets 0$ to \LW}{
        $\s \leftarrow$ \intPartition{i}\;
        \ForAll{$\bm{j}$ in \s}{
          $\e \leftarrow$ \new{$\bm{j}$,\ind}\;
          \If{$\bm{H} \cdot(\yhat \oplus \e)^\top == \bm{0}$} {
            $\estm \leftarrow (\yhat \oplus \e)\cdot\ginv$\;
            \KwRet{\estm}
            }
        }
    }
\end{algorithm}


\section{ORBGRAND Simplification}
\subsection{Generating the Integer Partitions}

An integer partition $\bm{\lambda}$ of a positive integer $m$, noted $\bm{\lambda} = (\lambda_1, \lambda_2, \ldots, \lambda_P) \vdash m$, where $\lambda_1>\lambda_2>\ldots>\lambda_P$, is the breakdown of $m$ into a sum of strictly positive integers $\lambda_i$. If all $\lambda_i$ are different, the partition is called distinct. In the ORBGRAND algorithm, each $\lambda_i$ represents an index to be flipped, and therefore it requires distinct integer partitions only. The generated test error pattern obtained from an integer partition with $P$ elements has a Hamming weight of $P$.

Hardware generation of the integer partitions was proposed in \cite{T14high-speedhardware}. However, their approach cannot be applied directly to our proposed ORBGRAND as the generated partitions are not distinct. In addition, their partition generation is sequential, which is not desirable for parallelized, high-throughput architectures. 
We noticed that when considering a specific logistic weight $LW$ convenient patterns appear. For example,
for $LW = 10$, the distinct integer partitions are $\bm{\lambda} = \{(10); (9,1); (8, 2); (7, 3);(6, 4); (7, 2, 1);(6, 3, 1);(5, 4, 1);$  $(5, 3, 2);(4, 3, 2, 1)\}$. If the listing order is followed for $P =2$ (i.e. the subset $\{(9,1); (8, 2); (7, 3);(6, 4)\}$ ), the first integer descends while the second ascends. Similar trends can be observed for higher-order partitions such as $P=3$: the first integer descends while the second ascends as the third integer remains fixed until all iterations for the first two integers are complete. Inspired from this trend, we propose an arrangement of shift registers to come up with a way of generating integer partitions of a particular size, that is described in Section \ref{sec:vlsi}.

Without loss of generality, $\forall i \in [2, P]$, when an integer is partitioned into $P$ parts, and assuming that $\lambda_i$ are 
ordered, the maximum value for each $\lambda_i$ is bounded by
\begin{equation}
 \lambda_i^\text{max} < \frac{2\times m - (i\times(i-1))+2-2\times\sum\limits_{j=i+1}^{P}\lambda_{j}}{2\times i}.
\label{eq:lambda_bound}    
\end{equation}
The first value of $\lambda$ can be found using $\lambda_1=m-\sum\limits_{j=2}^{P}\lambda_{j}$. The associated proof for (\ref{eq:lambda_bound}) is omitted due to the lack of space.

\subsection{Impact of the Parameters}

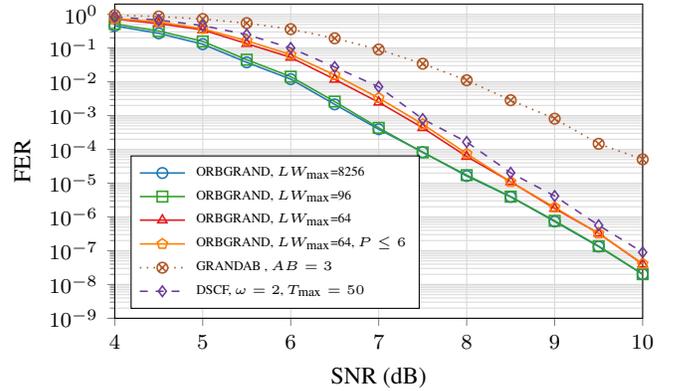
\begin{figure}[!t]
 \centering
\begin{tikzpicture}
    \begin{semilogyaxis}[ 
            footnotesize, width=\columnwidth, height=.67\columnwidth,    
            xmin=4, xmax=10, xtick={4,5,...,12},
            ymin=1e-9,  ymax=2,
            xlabel=SNR (dB), ylabel=FER,   
            grid=both, grid style={gray!30},
            tick align=outside, tickpos=left, 
            legend pos=south west, 
            legend cell align={left},
            /pgfplots/table/ignore chars={|},
            mark options={solid},
        ]

        \addplot[mark=o       , Paired-1 , semithick]  table[x=SNR, y=FER] {data/polar/N128_K105_C11_LW8256_HW128};
        \addplot[mark=square  , Paired-3 , semithick]  table[x=SNR, y=FER] {data/polar/N128_K105_C11_LW96_HW128}; 
        \addplot[mark=triangle, Paired-5 , semithick]  table[x=SNR, y=FER] {data/polar/N128_K105_C11_LW64_HW128}; 
        \addplot[mark=pentagon, Paired-7 , semithick]  table[x=SNR, y=FER] {data/polar/N128_K105_C11_LW64_HW6};
        \addplot[mark=otimes  , Paired-11, semithick, dotted]  table[x=SNR, y=FER] {data/polar/N128_K105_C11_ABGRAND3};
        \addplot[mark=diamond , Paired-9 , semithick, dashed]  table[x=snr_db, y=FER] {data/polar/quant/dscf};
        \legend{{} {{\tiny{ORBGRAND}, $LW_\text{max}$=8256}},
                {} {{\tiny{ORBGRAND}, $LW_\text{max}$=96}},
                {} {{\tiny{ORBGRAND}, $LW_\text{max}$=64}},
                {} {{\tiny{ORBGRAND}, $LW_\text{max}$=64, $P \leq 6$}},
                {} {{\tiny{GRANDAB }, $AB=3$}},
                {} {{\tiny{DSCF, $\omega=2$, $T_\text{max}=50$}}},
                }

    \end{semilogyaxis}
\end{tikzpicture}  
\caption{\label{fig:fer_polar}Comparison of decoding performance of PC(128,105+11) for ORBGRAND decoding with different
parameters, DSCF and GRANDAB.} 
\end{figure}

The maximum logistic weight ($LW_\text{max}$) has a strong impact on both the maximum number of codebook membership queries (and essentially, the worst-case complexity) and the decoding performance. Fig. \ref{fig:fer_polar} plots the frame error rate (FER) performance for ORBGRAND decoding of 5G CRC-aided polar code (128, 105+11) \cite{Duffy20205g,3GPP}, with BPSK modulation over an AWGN channel. The SNR in dB is defined as $\text{SNR} = -10\log_{10}\sigma^2$. For $LW_\text{max}$ values of 128, 96 and 64, the maximum number of required queries are $5.33\times10^{7}$, $3.69\times10^{6}$, and $1.5\times10^{5}$, respectively. When $LW_\text{max}$ is reduced from $128$ to $64$, a performance degradation of $0.2$ dB is observed at FER = $10^{-7}$. On the other hand, the complexity is reduced by $355\times$ as a result of this reduction. In addition to restricting the $LW_\text{max}$, the number of elements ($P$) in $\bm{\lambda}$ can also be limited. For example, for the considered polar code with $LW_\text{max}=64$, the degradation in error correction performance is negligible with $P=6$ when compared to an unlimited $P$. As a result of these simplifications, the maximum number of queries is reduced to $1.16\times10^{5}$. In addition to reducing complexity, appropriate selection of parameters $LW_\text{max}$ and $P$ helps design simple hardware implementations. 

In comparison with the hard decision variant of GRANDAB (AB=3), ORBGRAND results in a FER gain of at least 2 dB for 
FER lower than $10^{-5}$. Moreover, ORBGRAND with parameters $LW_\text{max}\leq64$ and $P\leq6$ results in a similar performance as the state-of-the-art Dynamic SC-Flip (DSCF) polar code decoder \cite{Chandesris}, which is also an iterative decoder. The number of attempts ($T_\text{max}$) parameter for DSCF is set to 50 and the maximum bit-flipping order $\omega$ is set to 2.

\section{VLSI Architecture for ORBGRAND}\label{sec:vlsi}
In \cite{GRANDAB-VLSI}, a VLSI architecture for GRANDAB (AB=3) was proposed. The VLSI architecture uses the linearity property of the underlying code to combine $t$ one bit flip error syndrome ($\bm{s}_i = \bm{H}\cdot\mathds{1}_i^\top$ with $i \in \llbracket 1\isep n \rrbracket$) to generate an error pattern with the Hamming weight of $t$. However, the VLSI architecture proposed in \cite{GRANDAB-VLSI} can only tackle up to 3-bit flips and the error patterns are tested in an ascending order of their Hamming weights. Thus, profound changes must be made to be able to consider soft-inputs, to generate error patterns in increasing order of their logistic weight, and to consider larger Hamming weights. 



\begin{figure}
  \centering
  \includegraphics[width=.65\linewidth]{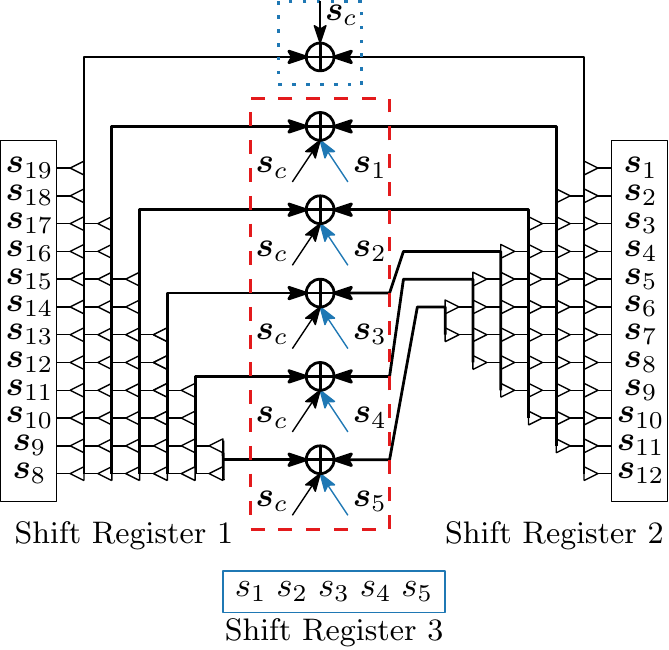}
  \caption{Example of the shift registers content and interconnection for $LW=20$.}
  \label{fig:registers} 
\end{figure}

\subsection{Principle, Scheduling and Details}
The VLSI architecture for the GRANDAB algorithm \cite{GRANDAB-VLSI} is articulated around shift registers storing syndromes of 1-bit flip error patterns. Since this approach allows us to move the data very slightly to compute different syndromes, we use this approach as the baseline for the proposed architecture for ORBGRAND. The number of shift registers has a direct impact on the Hamming weight of error patterns, that can be computed in parallel. However, if too many shift registers are used, the amount of interconnections become problematic. Therefore, we decide to consider 3 shift registers corresponding to an integer partition of size 3 ($P=3$).

During ORBGRAND decoding, for each $LW$ ($\forall LW \in [3,LW_\text{max}]$), integer partitions are generated with size $P$ ($\forall P \in [2,P_\text{max}]$). We propose to generate these integer partitions in ascending order of their size. This modification does not impact the FER performance, however, it helps for a simpler hardware implementation. In our proposed architecture, $\lambda_1$, $\lambda_2$, $\lambda_3$ are mapped to first, second and third shift register. The third shift register is a $\lambda_{3}^\text{max}\times(n-k)$ bit shift register, where $\lambda_{3}^\text{max}$ value is given by (\ref{eq:lambda_bound}). Whereas first and second shift registers have a size of $2\times(\lambda_{3}^\text{max}+1)\times(n-k)$ bits. Since we have $\lambda_1 = n-\sum_{i=2}^{3}\lambda_i$, $\bm{s_{n-i}}$ is stored at the $i$th index of the first shift register; whereas for the second and third shift registers $\bm{s_i}$ is stored at the $i$th index.

An example of the content and the interconnection of the three shift registers corresponding to $LW=20$ is presented in Fig. \ref{fig:registers}. The elements of the shift register 1 and 2 are connected to arrays of $(n-k)$-wide XOR gates whereby a collection of these connections is defined as a bus. The $\lambda_3^{max} + 1 = 6$ busses, each originating from a shift register, are used to feed the $(n-k)$-wide XOR gate arrays for computing the syndromes of the error patterns. The first array of XOR gates (dotted rectangle) is responsible to check for 2-bit flips error patterns by combining the syndromes of the received codewords with all the elements of the first and second register (the first bus of each shift register). Similarly, the remaining five XOR gate arrays combine the remaining buses ( the selected elements of the shift register 1 and 2) with the elements of the shift register 3 to check for 3-bit flips error patterns (dashed rectangle).

Due to the described layout of the 3 shift registers, all the error patterns corresponding to an integer partition of sizes 2 and 3 for a specific $LW$ are checked in one time-step. For generating test error patterns corresponding to integer partitions of sizes $P>3$, a controller is used in conjunction with the shift registers. The controller is able to combine $P_\text{max} - 3$ syndromes together, noted $\bm{s}_c$. Hence, when $\bm{s}_c$ is fixed, only one time-step is required to generate all possible combinations of $\{\lambda_1, \lambda_2, \lambda_3\}$ using the shift registers with adequately chosen shifting values. Therefore, the number of time steps required to generate all integer partitions of size $P>3$ for a specific $LW$ is given by:
\begin{equation}
\sum_{\lambda_P=1}^{\lambda_{P}^\text{max}}\left(\sum_{\lambda_{P-1}=\lambda_P+1}^{\lambda_{P-1}^\text{max}}\left(\ldots\sum_{\lambda_{4}=\lambda_5+1}^{\lambda_{4}^\text{max}}\left(1\right)\right)\right).  
\end{equation}

\begin{figure*}
\centering
  \includegraphics[width=0.85\linewidth]{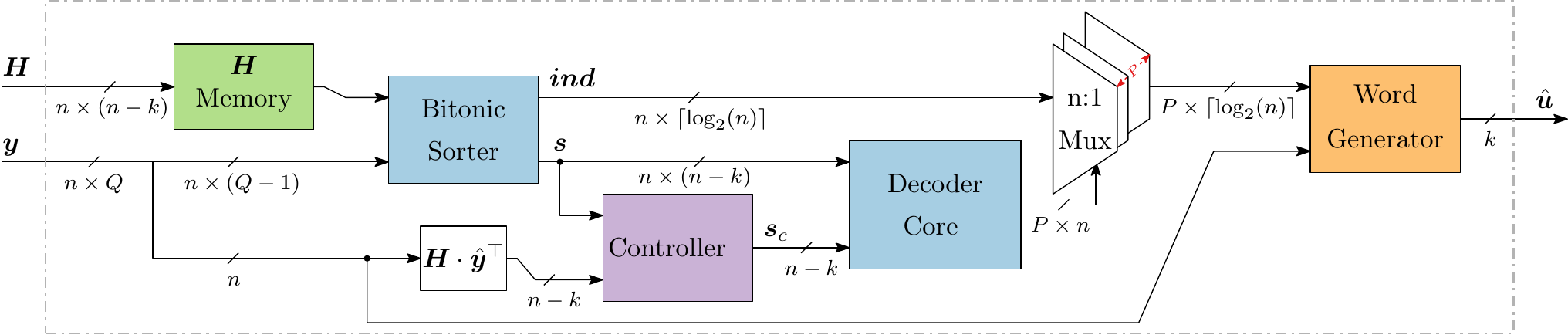}
  \caption{Proposed VLSI Architecture for ORBGRAND.}
  \label{fig:arch}
\vspace{-1em}
\end{figure*}

The proposed VLSI architecture for ORBGRAND is presented in Fig.~\ref{fig:arch}. The control and clock signals are not depicted for simplicity. Any $\bm{H}$ matrix can be loaded into the \textit{H memory} of size $n\times (n-k)\text{-bit}$ at any time to support different codes and rates. To achieve high throughput, a \textit{bitonic sorter} \cite{batcher68} is considered to sort the incoming LLRs.

At the first step of decoding, a syndrome check is performed on the hard decided word $\hat{\bm{y}}$. If the syndrome is verified, decoding is assumed to be successful. Otherwise, the bitonic sorter is invoked to sort the LLR values. The bitonic sorter is pipelined to $\log_2(n)$ stages and hence it takes $\log_2(n)$ cycles to sort the LLRs values. While indices of the sorted LLR are forwarded to the multiplexers for later use by the \textit{word generator} module, the sorted one bit-flip syndromes are passed to the \textit{decoder core}, as depicted in Fig.~\ref{fig:registers}. After the sorting operation, all the one bit-flip syndromes ($\bm{s_i}$,$\forall i \in [1,LW]$) are checked in one time-step. Next, error patterns are checked in the ascending logistic weight order. If any of the tested syndromes combinations verify the parity check constraint, a 2D priority encoder along with the \textit{controller} module is used to forward the respective indices to the word generator module, where $P$ multiplexers are used to translate the sorted index values to their correct bit flip locations.

\begin{table}[t]
\centering
\caption{\label{table:tableORBGRAND_128}TSMC 65 nm CMOS Implementation Comparison for ORBGRAND with GRANDAB and DSCF for $n=128$.}
\begin{adjustbox}{max width=\columnwidth}
\begin{tabular}{lrrr}
\toprule
                             & GRANDAB \cite{GRANDAB-VLSI} & ORBGRAND              & DSCF \cite{Ercan_2020}         \\
                             \cmidrule(l){2-2}\cmidrule(l){3-3}\cmidrule(l){4-4}
Parameters                   & AB=3    & LW$\leq$64, P$\leq$6 & $\omega = 2$, $T_\text{max}=50$ \\  
Technology (nm)              & 65      & 65                    & 65                              \\
Supply (V)                   & 0.9     & 0.9                   & 0.9                             \\
Max. Frequency (MHz)         & 500     & 454                   & 426                             \\
Area (mm\textsuperscript{2}) & 0.25    & 1.82                  & 0.22                            \\
W.C. Latency (ns)            & 8196    & 9308                  & 6103                            \\
Avg. Latency (ns)            & 2       & 2.47                  & 122                             \\
W.C. T/P (Mbps)              & 12.8    & 11.3                  & 17.2                            \\
Avg. T/P (Gbps)              & 52.5    & 42.5                  & 0.86                            \\
Code compatible              & Yes     & Yes                   & No                              \\
\bottomrule
\end{tabular}
\end{adjustbox}
\end{table}

\subsection{Implementation Results}
The proposed ORBGRAND architecture has been implemented in Verilog HDL and synthesized using the general-purpose TSMC 65 nm CMOS technology, using post-synthesis verified test vectors. Table \ref{table:tableORBGRAND_128} presents the synthesis results for the proposed decoder with $n=128$, code rates between $0.75$ and $1$, $LW\leq64$, $P\leq6$. Inputs are quantized on 5 bits, including 1 sign bit and 3 bits for the fractional part. 

The ORBGRAND implementation supports a maximum frequency of $454~\text{MHz}$. Since no pipelining strategy is used for the decoder core, one clock cycle corresponds to one time-step. For $n= 128$, $\numprint{4226}$ cycles are required in the worst-case (W.C.) scenario, resulting in a W.C. latency of 9.3$\mu$s. The average latency, however, is only of 2.47ns at SNR$=10$ dB (target FER of $10^{-7}$), which results in an average decoding information throughput of $42.5$ Gbps for a (128,105) 5G-NR CRC-Aided polar code. In comparison with the hard decision-based GRANDAB decoder (AB=3) \cite{GRANDAB-VLSI}, ORBGRAND has $7\times$ area overhead, and $13.5\%$ and $23\%$ higher W.C. and average latency. This translates into $13.3\%$ and $23.5\%$ lower W.C. and average decoding throughput, respectively. However, the FER performance of ORBGRAND, being a soft decision decoder, surpasses GRANDAB (AB=3) decoder by at least $2$ dB for target FER lower than $10^{-5}$, as shown in Fig. \ref{fig:fer_polar}.

The proposed ORBGRAND is also compared with the state-of-the-art iterative Dynamic SC-Flip (DSCF) polar code decoder with similar performance \cite{Ercan_2020}, considering a quantization of 6 and 7 bits for channel and internal LLRs, respectively. In comparison with DSCF, ORBGRAND has $8\times$ area overhead, in addition to $52\%$ overhead in worst-case latency. However, proposed ORBGRAND results in $49\times$ lower average latency and higher average throughput than the DSCF at a target FER of $10^{-7}$. Moreover, the proposed ORBGRAND is code and rate compatible, while DSCF can only decode polar codes.

\section{Conclusion}
In this paper, we presented the first high throughput hardware architecture for the ORBGRAND algorithm. Due to the code-agnostic nature of GRAND, the proposed architecture for ORBGRAND can decode any code, given the length and rate constraints. We suggest modifications in the ORBGRAND algorithm to assist hardware implementation as well as reducing the complexity.
ASIC synthesis results showed that an average decoding throughput of $42.5$ Gbps can be achieved for a code-length of $128$ for a target FER of $10^{-7}$. In comparison with the state-of-the-art DSCF decoder for polar codes, the proposed VLSI implementation results in $49\times$ decoding throughput for a 5G CA (128,105) polar code at an SNR of $10$ dB. Moreover, the proposed architecture serves as the first step towards the hardware implementation of soft-input decoders from GRAND family. 

\bibliographystyle{IEEEbib}
\bibliography{refs}

\begin{thebibliography}{10}

\bibitem{Hamming50}
R.~W. Hamming,
\newblock ``Error detecting and error correcting codes,''
\newblock {\em Bell System Technical Journal}, vol. 29, pp. 147--160, 1950.

\bibitem{turbo}
C.~{Berrou}, A.~{Glavieux}, and P.~{Thitimajshima},
\newblock ``Near shannon limit error-correcting coding and decoding:
  Turbo-codes. 1,''
\newblock in {\em Proceedings of ICC '93 - IEEE International Conference on
  Communications}, 1993, vol.~2, pp. 1064--1070 vol.2.

\bibitem{Gallager62}
R.~Gallager,
\newblock ``Low-density parity-check codes,''
\newblock {\em IRE Transactions on information theory}, vol. 8, no. 1, pp.
  21--28, 1962.

\bibitem{Arikan09}
E.~{Arikan},
\newblock ``Channel polarization: A method for constructing capacity-achieving
  codes for symmetric binary-input memoryless channels,''
\newblock {\em IEEE Transactions on Information Theory}, vol. 55, no. 7, pp.
  3051--3073, 2009.

\bibitem{Duffy19TIT}
K.~R. {Duffy}, J.~{Li}, and M.~{M{\'e}dard},
\newblock ``Capacity-achieving guessing random additive noise decoding,''
\newblock {\em IEEE Transactions on Information Theory}, vol. 65, no. 7, pp.
  4023--4040, 2019.

\bibitem{SRGRAND}
K.~R. {Duffy} and M.~{M{\'e}dard},
\newblock ``Guessing random additive noise decoding with soft detection symbol
  reliability information - {SGRAND},''
\newblock in {\em 2019 IEEE International Symposium on Information Theory
  (ISIT)}, 2019, pp. 480--484.

\bibitem{duffy2020ordered}
K.~R. Duffy,
\newblock ``Ordered reliability bits guessing random additive noise decoding,''
\newblock {\em arXiv preprint arXiv:2001.00546}, 2020.

\bibitem{Duffy20205g}
K.~R. Duffy, A.~Solomon, K.~M. Konwar, and M.~M{\'e}dard,
\newblock ``{5G NR CA-Polar} maximum likelihood decoding by {GRAND},''
\newblock in {\em 2020 54th Annual Conference on Information Sciences and
  Systems (CISS)}. IEEE, 2020, pp. 1--5.

\bibitem{T14high-speedhardware}
J.~Butler and T.~Sasao,
\newblock ``High-speed hardware partition generation,''
\newblock {\em ACM Transactions on Reconfigurable Technology and Systems}, vol.
  7, pp. 1--17, 01 2014.

\bibitem{3GPP}
{3GPP},
\newblock ``{NR; Multiplexing and Channel Coding},''
\newblock Tech. {R}ep. TS 38.212, April 2020,
\newblock {Rel. 16.1}.

\bibitem{Chandesris}
L.~{Chandesris}, V.~{Savin}, and D.~{Declercq},
\newblock ``Dynamic-scflip decoding of polar codes,''
\newblock {\em IEEE Transactions on Communications}, vol. 66, no. 6, pp.
  2333--2345, 2018.

\bibitem{GRANDAB-VLSI}
S.~M. {Abbas}, T.~{Tonnellier}, F.~{Ercan}, and W.~J. {Gross},
\newblock ``High-throughput {VLSI} architecture for {GRAND},''
\newblock in {\em 2020 IEEE Workshop on Signal Processing Systems (SiPS)},
  2020, pp. 1--6.

\bibitem{batcher68}
K.~E. Batcher,
\newblock ``Sorting networks and their applications,''
\newblock in {\em Proceedings of the April 30--May 2, 1968, Spring Joint
  Computer Conference}, New York, NY, USA, 1968, AFIPS '68 (Spring), p.
  307–314, Association for Computing Machinery.

\bibitem{Ercan_2020}
F.~{Ercan}, T.~{Tonnellier}, N.~{Doan}, and W.~J. {Gross},
\newblock ``Practical dynamic {SC}-flip polar decoders: Algorithm and
  implementation,''
\newblock {\em IEEE Transactions on Signal Processing}, vol. 68, pp.
  5441–5456, 2020.

\end{thebibliography}

\end{document}